\documentclass[journal]{IEEEtran}

\usepackage{float}
\usepackage{graphicx}
\usepackage{wrapfig}

\usepackage[labelfont=bf]{caption}
\usepackage{color,soul}
\usepackage{amsmath} 
\usepackage{mathtools} 
\usepackage{amsfonts} 
\usepackage{amssymb}

\usepackage{comment}

\usepackage{todonotes}

\usepackage{array}
\usepackage{booktabs}
\usepackage{rotating}
\usepackage{multirow}
\usepackage{multicol}
\usepackage{lscape}
\usepackage{subfig}

\usepackage{url}

\newcommand{\norm}[1]{\left\lVert#1\right\rVert}

\ifCLASSINFOpdf
\else
\fi
\begin{document}
%
\title{Image Evolution Trajectory Prediction and Classification from Baseline using Learning-based Patch Atlas Selection for Early Diagnosis}

%
%
%

\author{Can~Gafuro\u{g}lu,
        ~Islem~Rekik,~\IEEEmembership{Member,~IEEE}\thanks{Data used in preparation of this article were obtained from the Alzheimer's Disease Neuroimaging Initiative (ADNI) database (adni.loni.usc.edu). As such, the investigators within the ADNI contributed to the design and implementation of ADNI and/or provided data but did not participate in analysis or writing of this report. A complete listing of ADNI investigators can be found at: http://adni.loni.usc.edu/wp-content/uploads/how\_to\_apply/ADNI\_Acknowledgement\_List.pdf}
		
\thanks{C. Gafuro\u{g}lu and I. Rekik are affiliated with BASIRA Lab, School of Science and Engineering, University of Dundee, Dundee, UK. I. Rekik is co-affiliated with the Faculty of Computer and Informatics, Istanbul Technical University, Istanbul, Turkey}
\thanks{Corresponding author: Islem Rekik, \url{www.basira-lab.com}}} 

%
%

\markboth{IEEE Journal}%
{Gafuro\u{g}lu \MakeLowercase{\textit{et al.}}}
%



\maketitle

\begin{abstract}

Identifying Alzheimer's disease (AD) earlier before the neurodegeneration is too severe and where treatment is not currently available, might aid in preventing AD onset. Specifically, patients initially diagnosed with early mild cognitive impairment (eMCI) are known to be a clinically heterogeneous group with very subtle patterns of brain atrophy. To examine the boarders between normal controls (NC) and eMCI, Magnetic Resonance Imaging (MRI) was extensively used as a non-invasive imaging modality to pin-down subtle changes in brain images of MCI patients. However, despite the large body of research works on MCI/NC and major advances in neuroimaging technologies and brain image analysis and learning methods, eMCI research remains limited by the number of available MRI acquisition timepoints. These can be grouped into (i) single-timepoint and (ii) multi-timepoint (or longitudinal) based MCI diagnosis frameworks. Ideally, one would learn how to classify MCI patients with high accuracy from data acquired at a single timepoint, while leveraging `non-existing' follow-up observations. To this aim, we propose novel supervised and unsupervised frameworks that learn how to jointly predict and label the evolution trajectory of intensity patches, each seeded at a specific brain landmark, from a baseline intensity patch. Specifically, both strategies aim to identify the best training atlas patches at baseline timepoint to predict and classify the evolution trajectory of a given testing baseline patch. The supervised technique learns how to select the best atlas patches by training bidirectional mappings from the space of pairwise patch similarities to their corresponding prediction errors --when one patch was used to predict the other. On the other hand, the unsupervised technique learns a manifold of baseline atlas and testing patches using multiple kernels to well capture patch distributions at multiple scales. Once the best baseline atlas patches are selected, we retrieve their evolution trajectories and average them to predict the evolution trajectory of the testing baseline patch. Next, we input the predicted trajectories to an ensemble of linear classifiers, each trained at a specific landmark. Last, we use weighted majority voting to label the testing subject as NC or eMCI. Our classification accuracy increased by up to 10\% points in comparison to single timepoint-based classification methods.

\end{abstract}


%
\IEEEpeerreviewmaketitle

\section{Introduction}

%
%
%
%

\IEEEPARstart{D}{etecting} Alzheimer's Disease (AD) in a very early stage might help better monitor disease progression and improve the quality of lives of AD patients. Dementia, and AD by extension, is becoming an increasingly common problem as life expectancy goes up in developed countries \cite{Kalaria:2008}. 

Particularly, identifying early Mild Cognitive Impairment (eMCI) which is an early manifestation of Alzheimer's disease \cite{Morris:2001} remains a formidable challenge in dementia neuroscience. This is partly due to subtle anatomical fingerprint of eMCI, which makes it hard to differentiate from typical normal control (NC) brain anatomy and structure. However, disentangling the eMCI brain from the NC brain is a clinical problem with great significance due to the prevalence of AD \cite{Apostolova:2008}. Typically, neuroimaging and cognitive scores are widely used for AD diagnosis \cite{Frisoni:2010}. It is desirable to be able to diagnose dementia with only the use of structural Magnetic Resonance Imaging (MRI), as structural MRI scans can be taken quickly and at a low cost using equipment with widespread availability compared to other imaging modalities such as PET or functional MRI \cite{Coulthard:2017}.

Currently, this process places a burden on medical experts as they must individually examine structural MR images. Early MCI atrophy patterns are subtle and eMCI patients might show no additional clear signs of cognitive impairment aside from minor memory issues \cite{Jessen:2014}. Leveraging advanced machine learning can help automate eMCI diagnosis and alleviate the burden on these experts by providing them with reliable, automated and efficient reading and interpretation of MRI data. As such, a vast number of studies devised neuroimaging-based machine learning methods to predict and diagnose AD patients from a single MRI acquisition timepoint \cite{Magnin:2009, Kloppel:2008, Cuingnet:2011} or multiple available timepoints \cite{Sanroma:2017, Thung:2016, Zhu:2016, Zhang:2012}.

Among the longitudinal studies, \cite{Sanroma:2017} devised similarity maps capturing the relationship between registered baseline and follow-up neurimages to distinguish between stable MCI patients and MCI converters. \cite{Zhu:2016} designed a temporally structured support vector machine (SVM) classifier which captures the longitudinal unfolding of MCI over time by flexibly integrating any number of available follow-up MRIs to boost the classification performance. In another work, \cite{Thung:2016} demonstrated the potential of utilizing longitudinal neuroimaging data, even with incomplete measurements, to improve the classification of stable and converted MCI patients based on sparse modeling. A longitudinal multimodal MRI model was proposed in \cite{Zhang:2012} to learn how to predict MCI brain state evolution towards AD state by devising a sparse linear regression model with a group regularization constrained to group the weights corresponding to the same brain region across multiple time points so that selected altered brain regions are consistent across differnet timepoints. Next, by extracting longitudinal features from the original baseline and longitudinal data, an SVM is trained for classification. Notably, all these works all of these works incorporated multiple timepoints into their frameworks to leverage additional relevant information for increasing the classification accuracy of different demented brain states. However, these methods require more than a single acquisition timepoint for diagnosis. This might hinder the possibility of administering \emph{early} clinical treatment, if available, where early MCI patients are diagnosed with high accuracy \emph{solely} from baseline medical data. As potential preventive treatment \cite{Maliszewska:2017} is more likely to succeed the earlier the disease is detected, requiring subjects to wait for multiple measurements at different timepoints may impede their recovery.

\begin{figure*}[!ht]
\centering
\includegraphics[width=13cm]{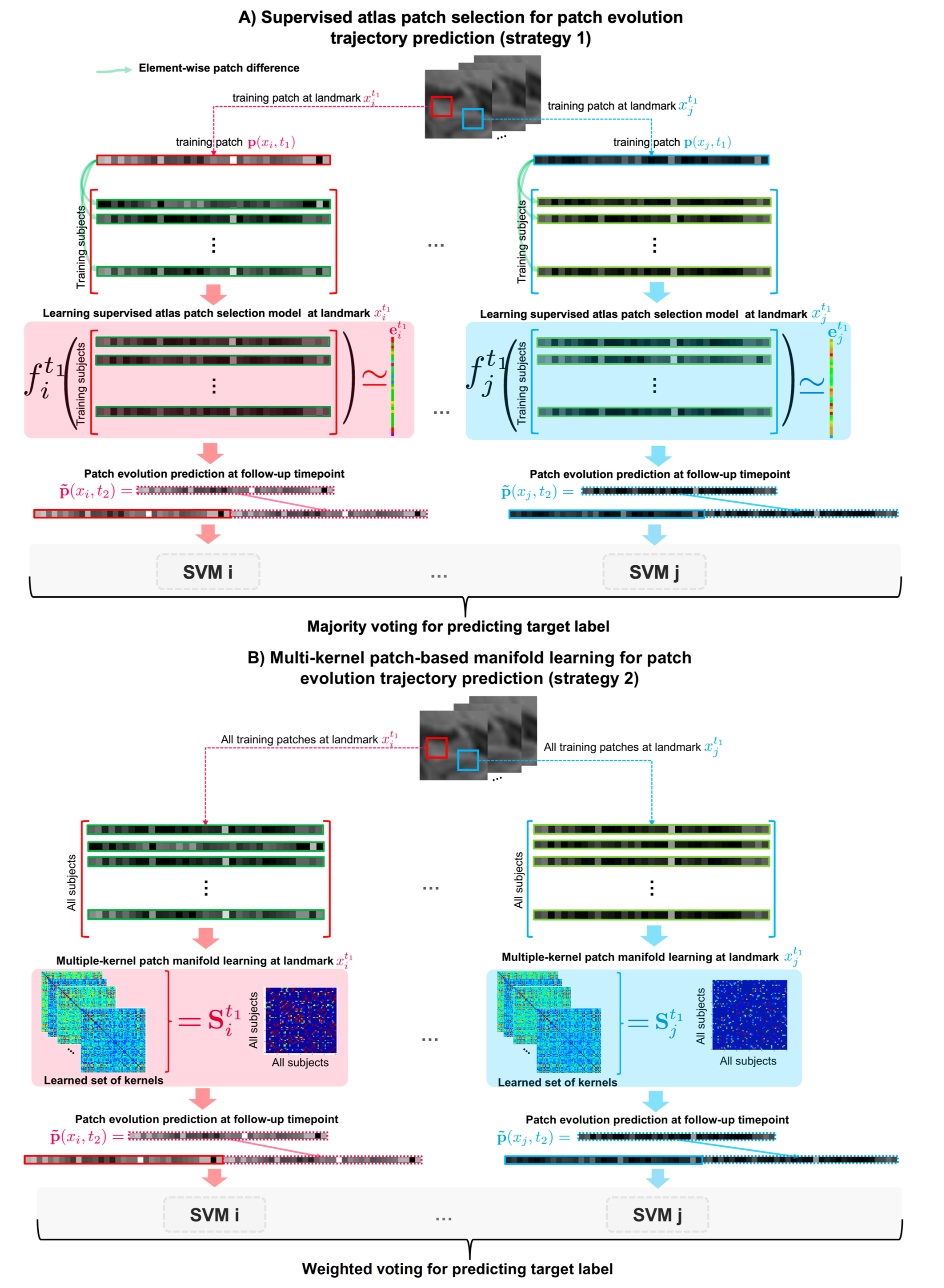}%
\caption{Illustration of the proposed patch-specific evolution trajectory prediction and classification framework from a baseline MRI using supervised (A) and unsupervised (B) strategies. (\textbf{A}) \emph{Supervised atlas patch selection strategy.} We first learn a mapping function $f_i^{t_1}$ at each landmark $x_i$, which aggregates two bidirectional regressors ${f^+}_i^{t_1}$ and ${f^-}_i^{t_1}$ (see Section II for more details), to map the intensity dissimilarity vector between a pair of patches to an error prediction vector. Basically, we compute the prediction error for a pair of training patches as the average of the prediction error produced when (i) using the first training patch to predict the evolution trajectory of the second training patch and (ii) using the second training patch to predict the evolution trajectory of the first training patch. This atlas patch selection strategy is supervised by the potential of the selected training atlases in producing low prediction errors at follow-up timepoint $t_2$. (\textbf{B}) \emph{Unsupervised atlas patch selection strategy.} We use multiple kernel learning to learn the pairwise similarity between training atlas patches and testing patch at baseline $t_1$, which can nicely capture the latent distributions of landmark-seeded patches at different bandwidths. This allows to identify to most similar baseline atlas patches to the target testing patch. Given the selected best atlas patches by strategy (\textbf{A}) or (\textbf{B}), we then linearly average the follow-up atlas patches at $t_2$ of the selected baseline atlas patches to predict the testing patch at $t_2$. Last, we train an ensemble of linear SVM classifiers to automatically label each predicted patch-wise evolution trajectory. By aggregating the predicted labels at each landmark using weighted majority voting, the label of the testing subject is predicted.}
\label{fig:1}
\end{figure*}

On the other hand, several other works focused on using a single acquisition timepoint for dementia diagnosis and classification, avoiding the limitation of requiring patients to wait for multiple scans. An ensemble of SVMs was proposed in \cite{Magnin:2009} to distinguish between NC and AD patients using MRI data. Ten classification methods from a single timepoint were compared in \cite{Cuingnet:2011} to perform three AD-related classification tasks: classifying AD against NC, MCI against NC, and MCI converters against stable MCI patients. In a different work, \cite{Kloppel:2008} proposed an automatic classification framework by training support vector machines to reliably distinguish AD from normal aging in individual structural MRI scans. 

However, these works missed out on the potential of incorporating longitudinal MRI data into their frameworks, which could further boost the classification performance by learning how to identify feature that best capture demented brain changes. More recently, a review paper \cite{Soussia:2018a} surveyed neuroimaging-based methods for dementia
diagnosis and prognosis published in MICCAI\footnote{International Conference on Medical Image Computing and Computer-Assisted Intervention} proceedings between 2010 and 2016. They identified 28 seed works developed using image or network brain data for MCI and AD diagnosis. Interestingly, predictive methods for early dementia diagnosis seem to be lagging behind, holding various untapped potentials for substantially advancing translational medicine.  Notably, the majority of the reviewed methods only focused on classification (e.g., NC vs MCI). While the ultimate goal of classification is to provide a computer-aided diagnosis for better clinical decisions, predicting future progression of early demented brains from a baseline observation (i.e., a single timepoint) remains a priority as it might help delay
conversion from MCI to AD when early treatment is addressed to the patient. For instance, if one can learn how to foresee abnormal local changes in the brain during MCI progression, these can provide more discriminative features that unfold over time for classifying MCI in a very early stage when such changes remain subtle compared with normal controls.

To overcome all these limitations, we propose to diagnose a patient at an \emph{early} stage solely based on a single baseline MRI data, leveraging longitudinal information that we \emph{learn how to accurately predict at follow-up timepoints}. Specifically, we  propose the first framework to jointly classify and predict the evolution trajectory of MRI from a single acquisition timepoint (i.e., baseline observation) in four key steps. First, we identify key voxels (or landmarks) at baseline $t_1$ in the target anatomical region of interest (ROI) across all training subjects. The detected landmarks seed the training of our methods as we aim to predict the evolution of intensity voxels in their local neighborhoods. Specifically, in the second step, we propose two novel \emph{supervised} and \emph{unsupervised} frameworks to predict patch evolution trajectory at each landmark, individually. Both strategies are rooted in the assumption that: \emph{if one can learn how to identify the best neighboring atlas patches to a given testing patch at baseline timepoint, one can use the available neighboring atlas patches at follow-up timepoints to predict the evolution trajectory of the testing patch over time.} Since the only available observation to predict from is at baseline timepoint, one can only examine the relationship between the testing samples and training samples at baseline to learn how to predict the missing follow-up observations. Simple but intuitive, such assumption showed great promise in learning how to predict the evolution of the multi-folded cortical surface from a single timepoint using neonatal MRI \cite{Rekik:2015,Rekik:2016a,Rekik:2017}. Given an input baseline testing patch, the first strategy learns how to select baseline `atlas patches' \emph{supervised} by their patch prediction error at a follow-up timepoint $t_2$. The second strategy leverages \emph{unsupervised} high-dimensional manifold using multiple-kernel learning to identify the local neighboring atlas patches to a given testing patch at a specific landmark. In the third step, we linearly average the follow-up atlas patches at $t_2$ of the selected baseline atlas patches to predict the testing patch at $t_2$. Last, we train an ensemble of linear SVM classifiers to automatically label each predicted patch-wise evolution trajectory. By aggregating the predicted labels at each landmark using weighted majority voting, the label of the testing subject is predicted.


\begin{table*}[h!]
\caption{ \emph{Major mathematical notations}.\label{tab:0}}
\centering
\begin{scriptsize}
\begin{tabular}{c@{~~}c@{~~}c}
	\toprule
	Mathematical notation & Definition \\
	\midrule
	$\mathbf{K}_l$ & $l$-th learning kernel in $\mathbb{R}^{n \times n}$ \\
	$n$ & number of subjects \\
	$\mathbf{S}$ & similarity matrix in $\mathbb{R}^{n \times n}$ for patch-based manifold learning \\
	$\mathbf{L}$ & latent matrix in $\mathbb{R}^{n \times c}$ \\
	$c$ & number of clusters \\
	$m$ & number of kernels \\
	$\mathbf{w}$ & weighting vector of the kernels in $\mathbb{R}^{m}$ \\
	$\mathbf{I}_n$ & identity matrix in $\mathbb{R}^{n \times n}$  \\
	$\mathbf{p}_{i,s}^{t_x}$ & patch centred on landmark $i$ for subject $s$ at timepoint $t_x$ \\
	$\mathbf{\tilde{p}}_{i,s}^{t_x}$ & predicted patch centred on landmark $i$ for subject $s$ at timepoint $t_x$ \\
	$\mathbf{\alpha}_{s, s'}$ & coefficient vector which maps the $t_1$ patch of subject $s'$ to the $t_1$ patch of subject $s$ \\
	${\mathbf{d}}_{s,s'}$ & element-wise difference between patches of subjects $s$ and $s'$ \\
	\bottomrule
\end{tabular}
\end{scriptsize}
\end{table*}

\begin{figure}[ht]
\centering
\includegraphics[width=8.5cm]{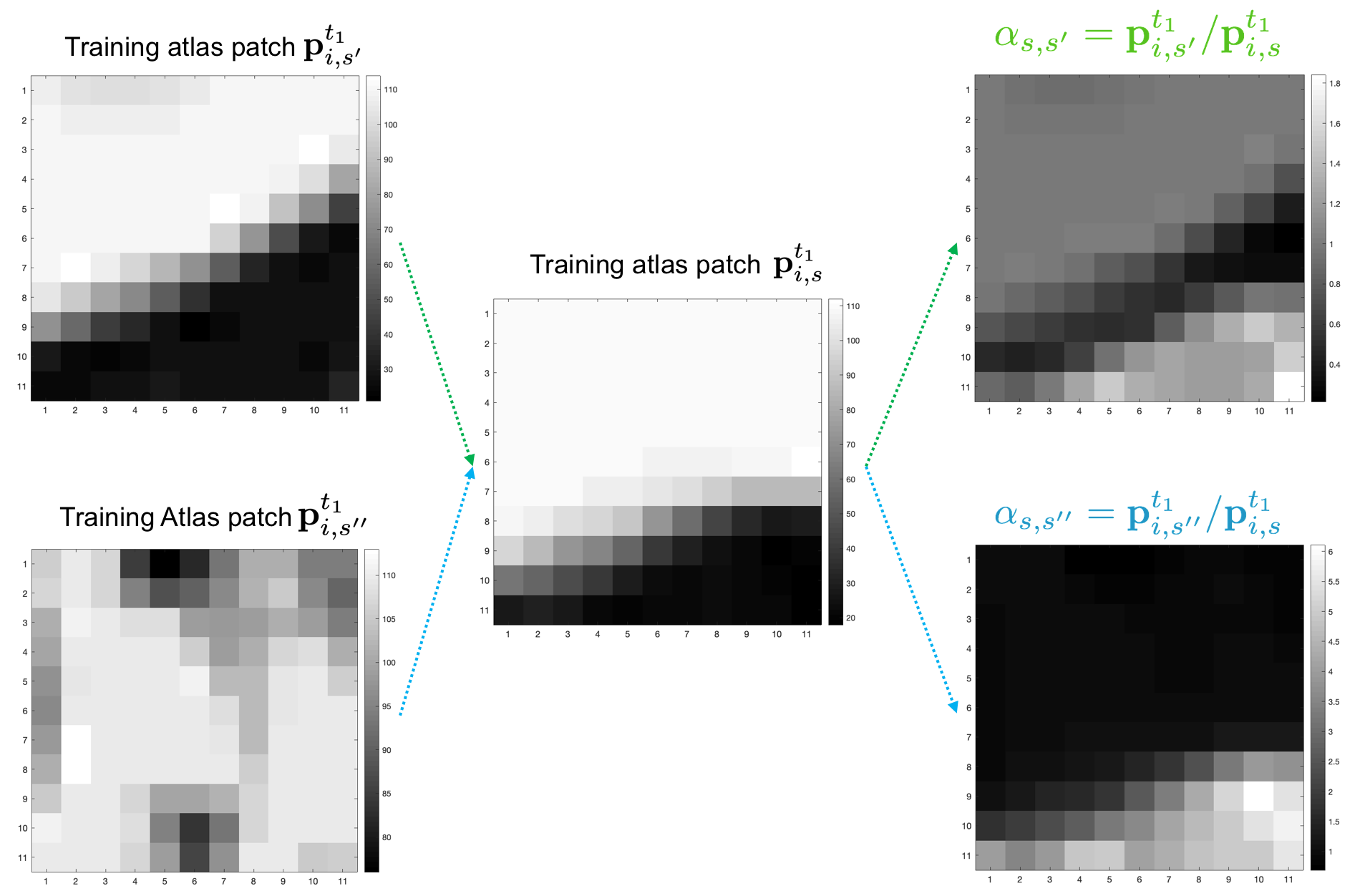}
\caption{\emph{Modeling the similarity between a baseline atlas patch $\mathbf{p}_{i,s'}^{t_1}$ of subject $s$ centered at landmark $i$ and a second baseline atlas patch $\mathbf{p}_{i,s}^{t_1}$ for the proposed supervised atlas selection (SAS) strategy.} We give an example of two training atlas patches with varying similarities to an input baseline training patch. We can clearly see how the quotient vector $\alpha_{s,s'}$ (resp., $\alpha_{s,s''}$) locally captures the degree of similarity between both patches.}
\label{fig:suppFig2}
\end{figure}

The contributions of our paper span multiple directions and present different kinds of advances:

\textbf{Conceptual advance.} To the best of our knowledge, this is the first work that aims to learn how to predict follow-up observations to better classify from a single baseline observation, with application to early MCI. Specifically, this paper defines the state-of-the-art in addressing this problem by proposing two innovative solutions based on supervised and unsupervised learning, which are rooted in the preliminary assumption that if one can learn how to identify the best atlas patches at baseline timepoint for the joint prediction and classification task, one can leverage the corresponding atlas patches at follow-up timepoints to predict and classify the evolution trajectory of a baseline testing patch.

\textbf{Technical advance.} We propose two novel techniques for learning how to select the best neighboring atlas patches for a given testing patch at baseline. The first one adapts multiple kernel manifold learning for building a manifold of baseline test and atlas patches to \emph{learn} how to model the relationship across training and testing patches using different bandwidths that can capture a heterogeneous distribution of patches at baseline. The second one proposes a novel strategy that trains bidirectional regressors to learn how to identify in a supervised manner the atlas patches, which if selected for predicting the testing patch evolution trajectory, would also lead to improving the classification accuracy of the testing subject as healthy or early demented.

\textbf{Clinical and translational advances.}  This is the first framework to jointly predict and classify brain image evolution over time from a single timepoint. It can be integrated into the clinical diagnosis framework for accurate brain disorder diagnosis and treatment planning using a single early observation. This will help alleviate the need to acquire multiple measurements (e.g., MRI scans) at different timepoints for diagnosis and prognosis. One early measurement will be sufficient given that the predictive model works well.
 
\textbf{Generic methodological advance.} The proposed core methods are generic and can be evaluated on any anatomical region of interest including the whole brain. They can be also utilized to improve the prediction accuracy by existing image and shape-based evolution prediction frameworks \cite{Rekik:2015,Rekik:2016a,Rekik:2017}, which simply used predefined metrics (such as Euclidean distance) for modeling the relationship between testing sample and atlases.\footnote{This research project won the Young Software Engineer of the Year in 2018: \url{https://digit.fyi/young-software-engineer-of-the-year-awards/} and \url{https://www.youtube.com/watch?v=6CcQCfnZrHM}.}


\section{Proposed Supervised and Unsupervised Patch-Based Evolution Trajectory Prediction and Classification from Baseline MRI}

Here, we detail the key steps of the proposed the supervised and unsupervised atlas patch selection strategies for jointly predicting and labeling landmark-seeded evolution trajectories while solely relying on a single MRI acquisition timepoint. 

\emph{Notations.} Throughout this paper, we denote matrices by boldface capital letters, e.g. $\mathbf{X}$, and scalars by lowercase letters, e.g. $x$. Bold lower case letter $\mathbf{x}$ stands for a vector. The transpose and is represented by $\mathbf{X}^T$. In addition, $\mathbf{I}$ denotes the identity matrix. For easy reference and enhancing the readability, we have summarised the major mathematical notations in \textbf{Table}~\ref{tab:0}.

\textbf{Fig.}~\ref{fig:1} displays the key steps of the proposed patch-specific evolution trajectory prediction and classification framework from a baseline MRI using supervised and unsupervised strategies. Both strategies share a common ground which charts the selection of the best training atlas patches for the target prediction and classification task, at each landmark individually. Using the supervised atlas patch selection strategy,  we first learn a mapping function $f_i^{t_1}$ at each landmark $x_i$, which aggregates two bidirectional regressors ${f^+}_i^{t_1}$ and ${f^-}_i^{t_1}$ (see Section II for more details), to map the intensity dissimilarity vector between a pair of patches to an error prediction vector. Basically, we compute the prediction error for a pair of training patches as the average of the prediction error produced when (i) using the first training patch to predict the evolution trajectory of the second training patch and (ii) using the second training patch to predict the evolution trajectory of the first training patch. This atlas patch selection strategy is supervised by the potential of the selected training atlases in producing low prediction errors at follow-up timepoint $t_2$. As for the unsupervised atlas patch selection strategy, we use multiple kernel learning to learn the pairwise similarity between training atlas patches and testing patch at baseline $t_1$, which can nicely capture the latent distributions of landmark-seeded patches at different bandwidths. This allows to identify to most similar baseline atlas patches to the target testing patch. Given the selected best atlas patches by either strategies, we then linearly average the follow-up atlas patches at $t_2$ of the selected baseline atlas patches to predict the testing patch at $t_2$. Last, we train an ensemble of linear SVM classifiers to automatically label each predicted patch-wise evolution trajectory. By aggregating the predicted labels at each landmark using weighted majority voting, the label of the testing subject is predicted. We detail below the steps of each proposed strategy.

\textbf{Landmark extraction.} Several brain disorders alter the morphology and structure of specific anatomical brain regions leading to local expansion or atrophy at the region boundary. Voxels located at the edge of target anatomical regions might present discriminative features to use for investigating the presence of eMCI in a particular brain. Hence, for each training subject, we apply a Sobel filter to the training label map (or segmentation image) of the target ROI across all slices to detect its edge. Next, by linearly averaging training edges of a particular ROI, we associate with each voxel and edge density value indicating its intensity occurence across training samples. Ultimately, by thresholding the edge density map, we identify the key training landmarks seeding the centers of our patches for devising the proposed supervised and unsupervised atlas patch selection strategies. For each ROI, we define the threshold as the mean minus the standard deviation of intensity distribution drawn from training patches. For a new testing subject, we extract patches centered at the landmarks set using the training samples.


\textbf{Supervised atlas patch selection for predicting patch evolution trajectory using bidirectional atlas patch to prediction error regressors.} Inspired by the work of \cite{Sanroma:2014} on learning how to select image atlases for accurate brain image segmentation, we propose a supervised atlas patch selection framework guided by the error a particular base line atlas at $t_1$  can produce when selected for predicting the patch evolution trajectory of a testing subject at a follow-up timepoint $t_k,\ \ \ k\geq2$. Specifically, given $n-1$ training patches, we train our supervised atlas selection model for each left out testing subject. To do so, for each pair of training patches in the training set comprising $n-1$ patches, we compute their element-wise absolute difference as shown in \textbf{Fig.}\ref{fig:1}--A. While excluding self-differences, this produces an \emph{intensity disparity matrix} of size $((n-1) \times (n-2))$ at $t_1$, where each row ${\mathbf{d}}_{s,s'} = |\mathbf{p}_{i,s}^{t_1} - \mathbf{p}_{i,s'}^{t_1}|$ denotes an element-wise absolute difference between two training patches $\mathbf{p}_{i,s}^{t_1}$ and $\mathbf{p}_{i,s'}^{t_1}$.

\begin{figure}[ht]
\centering
\includegraphics[width=8.5cm]{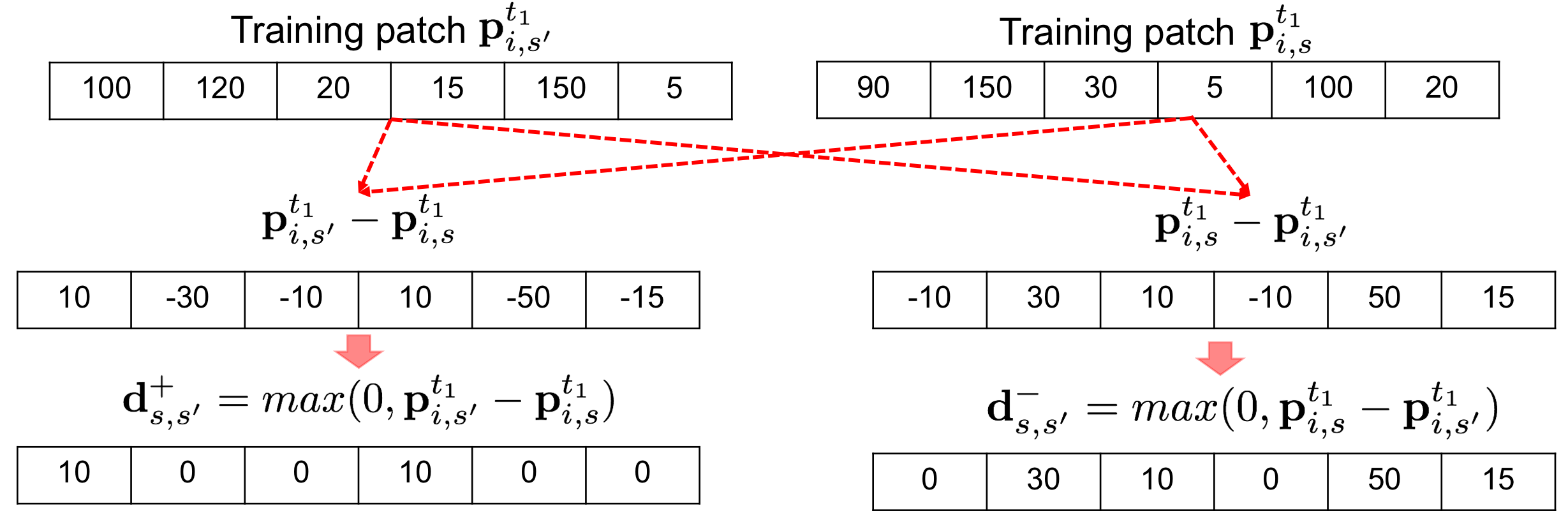}
\caption{\emph{Positive and negative disparity matrices construction.} Given two training baseline patches $\{ \mathbf{p}_{i,s}^{t_1}, \mathbf{p}_{i,s'}^{t_1}  \}$, both centered at landmark $i$ acquired at baseline timepoint $t_1$ for different subjects $s$ and $s'$, we compute their element-wise bi-directional differences:  $\mathbf{d}_{s,s'}^{+}$ and $\mathbf{d}_{s,s'}^{-}$, each representing a row vector in the positive and negative disparity matrices, respectively. In this figure, we give an example of vectorized patches.}
\label{fig:suppDisp}
\end{figure}

Next, we build a corresponding error vector $\mathbf{e}_i^{t_1}$ that quantifies the prediction error of using $\mathbf{p}_{i,s}^{t_1}$ for subject $s$ at $t_1$ as an atlas patch to predict $\mathbf{\tilde{p}}_{i,s^\prime}^{t_2}$ for subject $s^\prime$ at $t_2$. Specifically, we define $\mathbf{\tilde{p}}_{i,s^\prime}^{t_2}=\mathbf{\alpha}_{s,s'} \times\mathbf{p}_{i,s}^{t_2}$, where $\mathbf{\alpha}_{s,s'}$ is a vector of coefficients defined by $\mathbf{\alpha}_{s,s'}=\ \mathbf{p}_{i,s^{\prime}}^{t_1}\ /\ \mathbf{p}_{i,s}^{t_1}$ (\textbf{Fig.}~\ref{fig:suppFig2}). Basically, $\mathbf{\alpha}_{s,s'}$ maps the $t_1$ patch of subject $s'$ to the $t_1$ patch of subject $s$.  If $\mathbf{p}_{i,s}^{t_1}$ has a zero-element, then the corresponding element in vector $\mathbf{\alpha}_{s,s'}$ is set to a high value. We note that we chose the intensity quotient of two patches $\mathbf{p}$ and $\mathbf{q}$ as it allows to model the bi-directional similarity ($\mathbf{p}/\mathbf{q} \ \neq \mathbf{q}/\mathbf{p}$) between two patches as opposed to using the absolute distance ($|\mathbf{p} - \mathbf{q}| = |\mathbf{q} - \mathbf{p}|$). It also avoids producing negative similarity values as opposed to computing the simple difference ($\mathbf{p} - \mathbf{q}$). Next, we define the prediction error as the average absolute difference between the ground truth patch $\mathbf{p}_{i,s^\prime}^{t_2}$ and the predicted patch $\mathbf{\tilde{p}}_{i,s^\prime}^{t_2}$ for the testing subject $s^{\prime}$. In the training stage, we learn a support vector regressor function $f_i^{t_1}$ that maps the intensity disparity matrix at each landmark $x_i$ onto the corresponding prediction error vector $\mathbf{e}_{i}^{t_1}$ (\textbf{Fig.}~\ref{fig:1}--A). 

However, we note that if a subject $a$ is used to predict another subject $b$ at $t_2$, the prediction error $\norm{\mathbf{p}_{i,b}^{t_2}- \mathbf{\tilde{p}}_{i,b}^{t_2}}_2$ is likely to differ from the prediction error $\norm{\mathbf{p}_{i,a}^{t_2}-\mathbf{\tilde{p}}_{i,a}^{t_2}}_2$ obtained when subject $b$ is used to predict subject $a$ at $t_2$. However, the magnitude$\ m=\norm{d_{a,b}}_2= \norm{d_{b,a}}_2$ of the two disparity matrices $d_{a,b}$ and $d_{b,a}$ are the same, hence the support vector regressor function ends up with a single aggregated disparity value mapping to multiple values of prediction errors for every training subject pair. In order to avoid this, we propose to cleave the intensity disparity matrix into two matrices accounting for the direction in which the difference is computed: (i) one from $\mathbf{p}_{i,s'}^{t_1}$ to $\mathbf{p}_{i,s}^{t_1}$, and (ii) one from $\mathbf{p}_{i,s}^{t_1}$ to $\mathbf{p}_{i,s'}^{t_1}$ as illustrated in \textbf{Fig.}~\ref{fig:suppDisp}. Basically, in the `positive' unidirectional intensity disparity matrix, each row denoting the distance between two patches $\mathbf{p}_{i,s}^{t_1}$ and $\mathbf{p}_{i,s'}^{t_1}$ is defined as $\mathbf{d}^{+}_{s,s'} = max(0, \mathbf{p}_{i,s'}^{t_1}-\mathbf{p}_{i,s}^{t_1})$ (resp. $\mathbf{d}^{-}_{s,s'} = max(0, \mathbf{p}_{i,s}^{t_1}-\mathbf{p}_{i,s'}^{t_1})$ in the `negative' unidirectional intensity disparity matrix). To distinguish between both unidirectional disparity matrices, we use `positive' and `negative' directional differences, however, elements in both matrices are positive (\textbf{Fig.}~\ref{fig:suppDisp}). 
In the next step, we train two separate regressors: one regressor function ${f^+}_i^{t_1}$ on the positive disparity matrix and another regressor function ${f^-}_i^{t_1}$ on the negative disparity matrix. Since $d_{a,b}\neq\ d_{b,a}$ unless $a\ =\ b$, each regressor function ${f^-}_i^{t_1}$ and ${f^+}_i^{t_1}$ maps a unique disparity aggregate value to a unique prediction output. The output of the function $f_i^{t_1}$ predicting the overall prediction error given the intensity disparity matrix is then defined as $f_i^{t_1}= ({f^+}_i^{t_1} + {f^-}_i^{t_1})/2$.
In the testing stage, we compute the pairwise positive distance $\mathbf{d}^{+}_{i,tst}$ between each training patch $\{ \mathbf{p}_{i,s}^{t_1} \}_{s=1}^{n-1}$ and the testing patch $\mathbf{p}_{i,tst}^{t_1}$. Then, by testing the learned regression function ${f^+}_i^{t_1}$ for each pairwise distance ${f^+}_i^{t_1}(\mathbf{d}^{+}_{i,tst})$, we predict the error of using subject $s$ to predict $\mathbf{p}_{i,tst}^{t_2}$. Similarly, we compute the pairwise absolute `negative' distance $\mathbf{d}^{-}_{i,tst}$ between each training patch $\{ \mathbf{p}_{i,s}^{t_1} \}_{s=1}^{n-1}$ and the testing patch $\mathbf{p}_{i,tst}^{t_1}$. Next, we test the learned regression function ${f^-}_i^{t_1}$ to predict the error of using subject $s$ to predict $\mathbf{p}_{i,tst}^{t_2}$. Ultimately, each training atlas patch will be assigned the average error $({f^+}_i^{t_1}(\mathbf{d}^{+}_{i,tst}) + {f^-}_i^{t_1}(\mathbf{d}^{-}_{i,tst})/2)$. Ultimately, we select the top K atlas patches at $t_1$ with the lowest prediction errors, then average their corresponding patches at $t_2$ to output $\mathbf{\tilde{p}}_{i,tst}^{t_2}$.

\textbf{Unsupervised patch selection for predicting patch evolution trajectory using multi-kernel patch manifold learning.} To identify the baseline atlas patches whose follow-up images best represent the baseline testing patch in an unsupervised manner, we propose to learn pairwise patch intensity similarities (\textbf{Fig.}~\ref{fig:1}--B). Fundamentally, we adapt the MKML framework introduced in \cite{Wang:2017} to cluster generic data to our aim. MKML efficiently learns sample-to-sample similarity measure that best fits the structure of the data by combining multiple Gaussian kernels. It demonstrated significant outperformance in comparison with clustering methods that used pre-defined similarity measures such as Euclidean similarity and Pearson correlation, instead of learning it in a data-driven manner \cite{Wang:2017}. Given $n$ samples, each represented by a  $d$-dimensional feature vector (i.e., patch intensities in our case), MKML outputs an $n \times n$  patch similarity matrix $\mathbf{S}$.

Given a testing baseline patch $\mathbf{\tilde{p}}_{k,tst}^{t_1}$ seeded at landmark $x_k$, we first learn a baseline similarity matrix $\mathbf{S}_k^{t_1}$ modeling the relationship between baseline training and testing patches. Instead of using one predefined distance measure which may fail to capture the nonlinear relationship in the patch data, we use $m$ multiple Gaussian kernels $\{ \mathbf{K}_l \}_{l=1}^{m}$, weighed by a set of coefficients $\{ w_l \}_{l=1}^{m}$ to learn. In fact, by estimating the weights associated with each kernel, one can model the similarity between patches at different scales, thereby capturing spread out patch distributions as well as dense patch distributions when present in the dataset of interest. Adopting multiple kernels allows to better fit the true underlying statistical distribution of intensity patches. Besides, multiple kernels have been shown to correspond to different informative representations of the data and often are more flexible than a single kernel \cite{Wang:2017, Gonen:2011}.

Additionally, constraints are imposed on kernel weights to avoid a single kernel selection \cite{Wang:2017}. 

The Gaussian kernel at scale $\sigma_l$ is expressed as follows: $\mathbf{K}_l(\mathbf{p}_{i,s}^{t_1},\mathbf{p}_{i,s'}^{t_1}) = \frac{1}{\epsilon_{s,s'} \sqrt{2 \pi}} e^{ (- \frac{|\mathbf{p}_{i,s}^{t_1} - \mathbf{p}_{i,s'}^{t_1}|^2}{2 \epsilon_{s,s'}^2}) }$, where  $\epsilon_{s,s'}$ is defined as: $\epsilon_{s,s'} = \sigma_l(\mu_s + \mu_{s'})/2$, where $\sigma_l$ is a tuning parameter and $\mu_s = \frac{ \sum_{l \in KNN(\mathbf{p}_{i,s}^{t_1})} |\mathbf{p}_{i,s}^{t_1} - \mathbf{p}_{i,s'}^{t_1}| }{k}$, where $KNN(\mathbf{p}_{i,s}^{t_1})$ represents the top $k$ neighboring subjects of subject $s$. The computed kernels are then averaged to further learn the similarity matrix $\mathbf{S}_k^{t_1}$ at landmark $x_k$ and baseline timepoint $t_1$ through an optimization framework formulated as follows:

$\min_{\mathbf{S}_k^{t_1},\mathbf{L},\mathbf{w}} \Big[ \sum_{i,j,l} - w_l \mathbf{K}_l(\mathbf{h}^i,\mathbf{h}^j) \mathbf{S}{_k^{t_1}}(i,j) + \beta ||\mathbf{S}_k^{t_1}||_F^2 + \gamma \textbf{tr}(\mathbf{L}^T (\mathbf{I}_n - \mathbf{S}_k^{t_1}) \mathbf{L}) + \rho \sum_l w_l log w_l \Big]$

Subject to: $\sum_l w_l =1$, $w_l \geq 0$, $\mathbf{L}^T \mathbf{L}  = \mathbf{I}_c$, $\sum_j \mathbf{S}{_k^{t_1}}(i,j) = 1$, and $\mathbf{S}{_k^{t_1}}(i,j) \geq 0$ for all $(i,j)$, where:

$\bullet$ $\sum_{i,j} - w_l \mathbf{K}_l(\mathbf{h}^i,\mathbf{h}^j) \mathbf{S}{_k^{t_1}}(i,j)$ refers to the relation between the similarity and the kernel distance with weights $w_l$ between two subject-specific patches. The learned similarity should be small if the distance between a pair of patches is large.
	
$\bullet$ $\beta ||\mathbf{S}_k^{t_1}||_F^2$ denotes a regularization term that avoids over-fitting the model to the data.
	
$\bullet$ $\gamma \textbf{tr}(\mathbf{L}^T (\mathbf{I}_n - \mathbf{S}_k^{t_1}) \mathbf{L})$: $\mathbf{L}$ is the latent matrix of size $n \times c$ where $n$ is the number of subjects and $c$ is the number of clusters. The matrix $(\mathbf{I}_n -\mathbf{S}_k^{t_1})$ denotes the graph Laplacian.  
	
$\bullet$ $\rho \sum_l w_l log w_l$ imposes constraints on the kernel weights to avoid selection of a single kernel.

An alternating convex optimization is adopted where each variable is optimized while fixing the other variables until convergence \cite{Wang:2017}. Finally, based on the landmark-specific learned matrix $\mathbf{S}_k^{t_1}$, we select the top $K$ training patches (or $K$ atlas patches) that are most similar to the target testing patch at baseline. Finally, we predict $\mathbf{\tilde{p}}_{i,tst}^{t_2}$ as a weighted average of corresponding $K$ atlas patches at follow-up $t_2$.

\textbf{Predicted patch-based trajectory labeling using ensemble SVM classifier}. Last, at each landmark $x_i$, we classify the patch evolution trajectory for a testing subject as `healthy' or `disordered'. Given a landmark-seeded predicted patch evolution trajectory, we train a linear SVM classifier using the concatenation of baseline training patches $\{ \mathbf{p}_{i,s}^{t_1} \}_{s=1}^{n-1}$ and their corresponding predicted patches at follow-up timepoints $\{ \mathbf{\tilde{p}}_{i,s}^{t_2} \}_{s=1}^{n-1}$ obtained using the strategies detailed previously. The left out testing subject is then labelled using a weighted voting scheme on predicted labels outputted by all SVMs (i.e. across all landmarks). The weights of the votes are assigned based on the posterior probabilities of the classification being correct, as estimated by the SVM classifiers.

\section{Results}

\textbf{Data and model parameters.} We evaluated both supervised and unsupervised strategies using 30 NC (Normal Control) and 30 eMCI subjects acquired from the Alzheimer's Disease Neuroimaging Initiative (ADNI GO) \cite{Mueller:2005,Jack:2008,Aisen:2010} database (\url{adni.loni.usc.edu}), each with two T1-weighted MRIs (a baseline and a 6-month follow-up). The ADNI was launched in 2003 as a public-private partnership, led by Principal Investigator Michael W. Weiner, MD. The primary goal of ADNI has been to test whether serial magnetic resonance imaging (MRI), positron emission tomography (PET), other biological markers, and clinical and neuropsychological assessment can be combined to measure the progression of mild cognitive impairment (MCI) and early Alzheimer's disease (AD). 

Each MRI was pre-processed using the following steps as detailed in \url{http://adni.loni.usc.edu/methods/mri-tool/mri-analysis/} correction steps. These include (1) correction of image geometry distortion due to gradient non-linearity using Gradwarp \cite{Tan:2013}, (2) B1 non-uniformity correction of image intensity, and (3) N3 intensity correction using  histogram peak sharpening algorithm to reduce residual image intensity non-uniformity \cite{Boyes:2008}. All T1-weighted images have a resolution of $1.2 \times 1.01 \times 1.01$mm and a volume size of $196 \times 256 \times 256$. Next, we fed each T1-weighted image into FreeSurfer \cite{Fischl:2012} which performs skull stripping, brain mask extraction, brain tissue segmentation into different anatomical regions of interest by registration to the Montreal Neurological
Institute (MNI) space \cite{Collins:1995}. Each registered and segmented T1-weighted image is sampled to a uniform $1^3$mm resolution and a volume of size $256 \times 256 \times 256$.

For evaluation, we tested each method using a leave-one-out cross-validation scheme on landmark patches extracted from the left and right hippocampi and the lateral ventricles. We selected these preliminary ROIs due to their prevalence in the dementia literature \cite{Cuingnet:2011,Sanroma:2017} as well as the demonstrated link between AD and the atrophy rates of the hippocampus \cite{Morra:2009, Apostolova:2006,Deweer:1995, deLeon:1993} as well as the expansion of the lateral ventricles  \cite{Thompson:2004,Nestor:2008}. 

Hyperparameter optimization is used to tune and set each linear SVM classifier's cost parameter $C$, using a nested cross-validation scheme with 5 folds on the training data. We evaluate values from an exponentially growing sequence between $2^{-6}$ and $2^{15}$ to find the best assignment for $C$. We fixed the patch size to $11 \times 11 \times 11$ across all methods and ROIs. For MKML, we first used grid search to optimize both the number of clusters $c$ and the number of kernels across all landmarks. For the number of clusters, we explored the range between 1 and 3. The optimal number of clusters generally was set to 2 or 3 clusters since when exceeding 3 clusters, the patch clusters become imbalanced. $c= \{ 2,3 \}$ captures well the baseline patch distribution across landmarks. For MKML parameters, we set the number of clusters $c=2$ for the right hippocampus and set $c=3$ for all other regions. As for the number of kernels, we explored values between 1 and 10. Next, for each strategy, we optimized the third hyperparameter $K$. Basically, by varying  the number of atlas patches $K$ between 1 and 6, we noted that the performance starts dropping when generally exceeding 2 atlas patches. Hence, we ended up selecting 1 or 2 depending on the target ROI.  Specifically, across all ROIs, MKML strategy was run with the parameter $K$ (the number of atlas patches selected for prediction) set to $1$. For the supervised atlas selection (SAS) strategy, $K$ was set to 1 for the left hippocampus, $2$ for the right hippocampus and the left lateral ventricle and $3$ for the right lateral ventricle. For MKML, we set the number of kernels to $3$ for the left and right lateral ventricles, $5$ for the right hippocampus, and $7$ for the left hippocampus. The edge density threshold for choosing landmark voxels has been set to $0.19$ for the left lateral ventricle, $0.18$ for the right lateral ventricle, and $0.15$ for the hippocampi based on the automatically defined threshold using the mean and standard deviation of the intensity distribution in each target ROI.

\begin{figure*}
\centering
\vspace{-12pt}
\includegraphics[width=11cm]{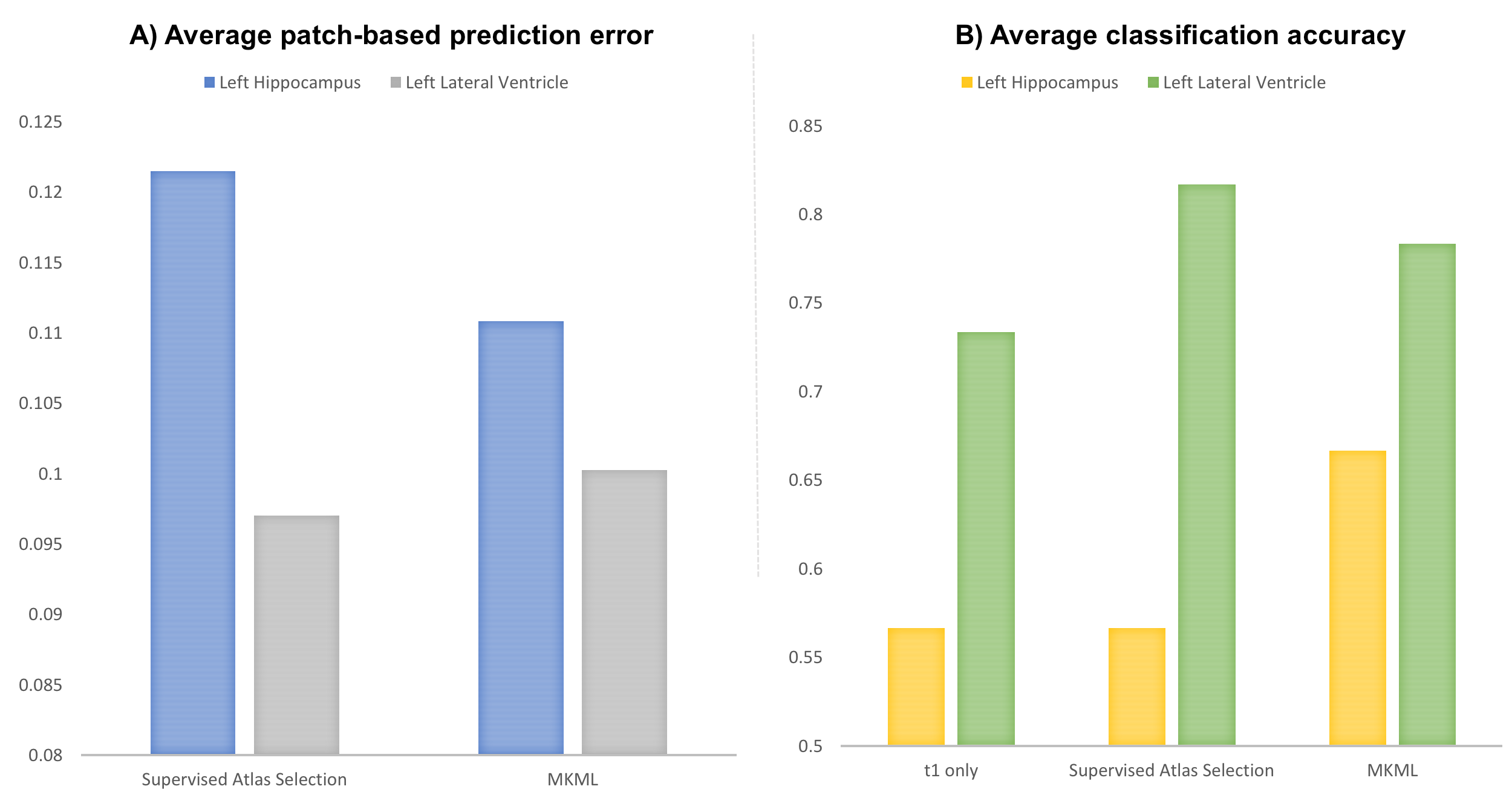}
\includegraphics[width=11cm]{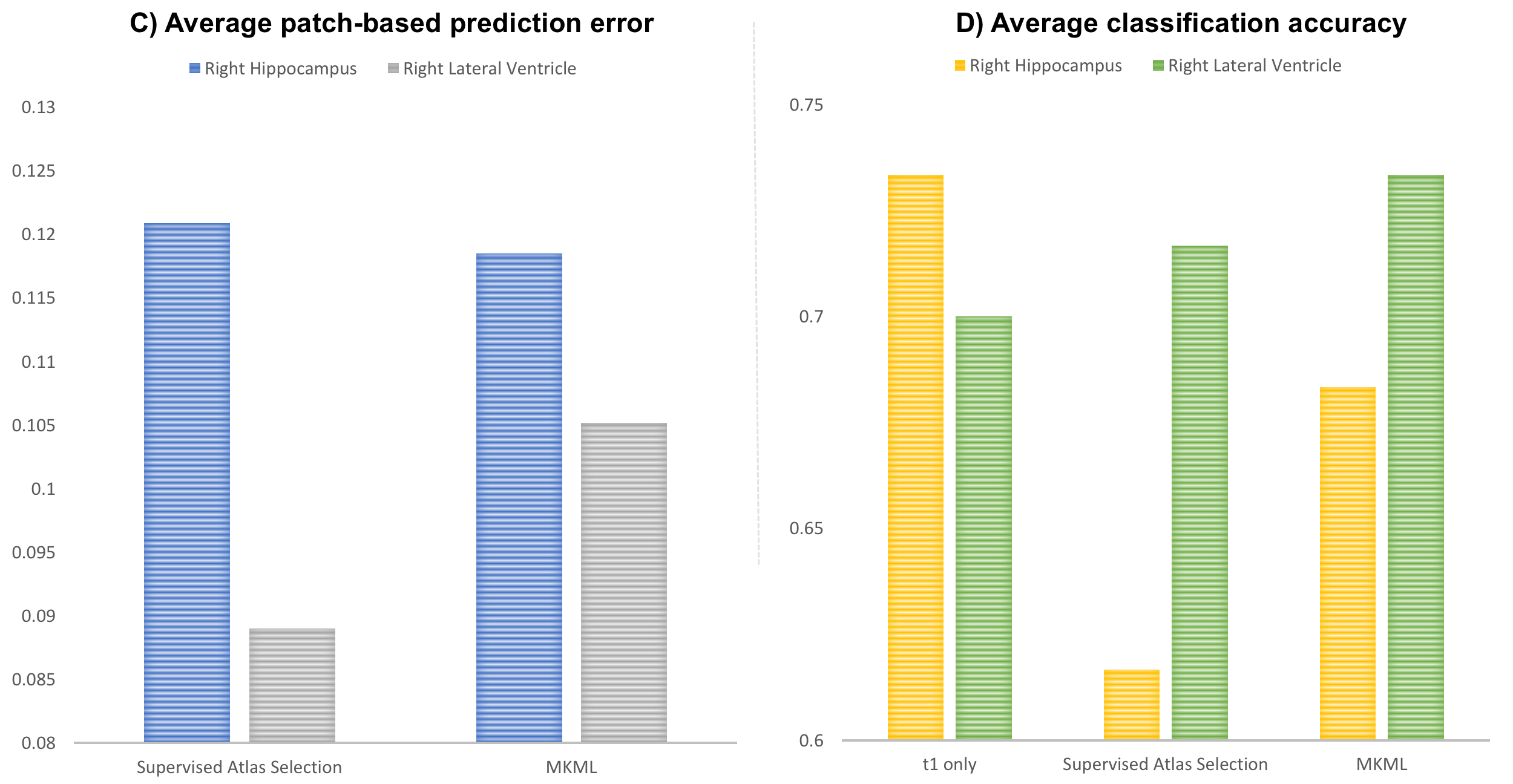}
\vspace{-6pt}
\caption{\emph{Image evolution trajectory prediction and classification results.} (A) and (C): Average predicted patch accuracy evaluated using mean absolute error (MAE) for each ROI in the left and right hemisphere respectively, using the proposed strategies for predicting the follow-up image evolution trajectory from baseline. (B) and (D): The average classification accuracy of our proposed methods for each ROI in the left and right hemisphere, respectively.}
\label{fig:4}
\end{figure*}

\begin{table*}
\centering
\caption{ NC/eMCI classification performance using the proposed strategies and baseline-only classification.}
\includegraphics[width=18cm]{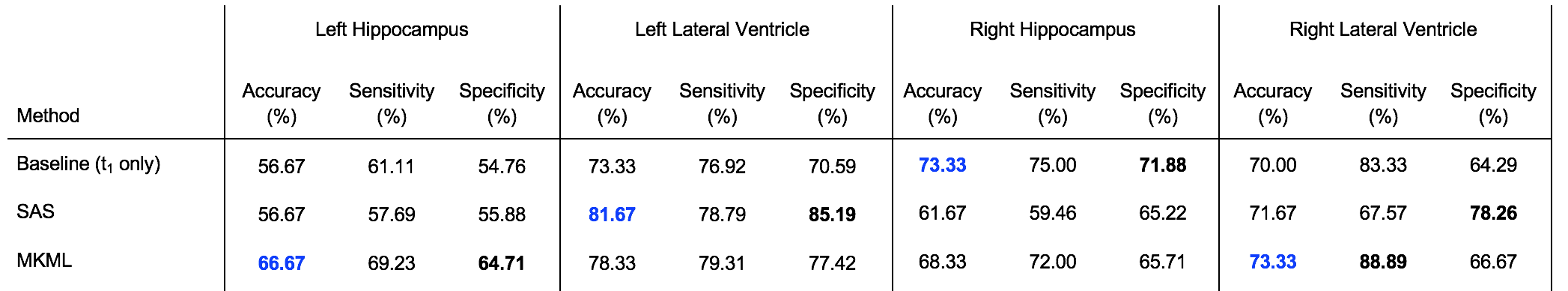}
\vspace{-10pt}
\label{tab:1}
\end{table*}

\begin{table}
\centering
\caption{Prediction accuracy for follow-up timepoint using average Pearson correlation between the predicted patches and the ground truth patches in four regions of interest. (LH):  left hippocampus. (LV): left ventricle. (RH): right hippocampus. (RV): right ventricle.}
\includegraphics[width=8.5cm]{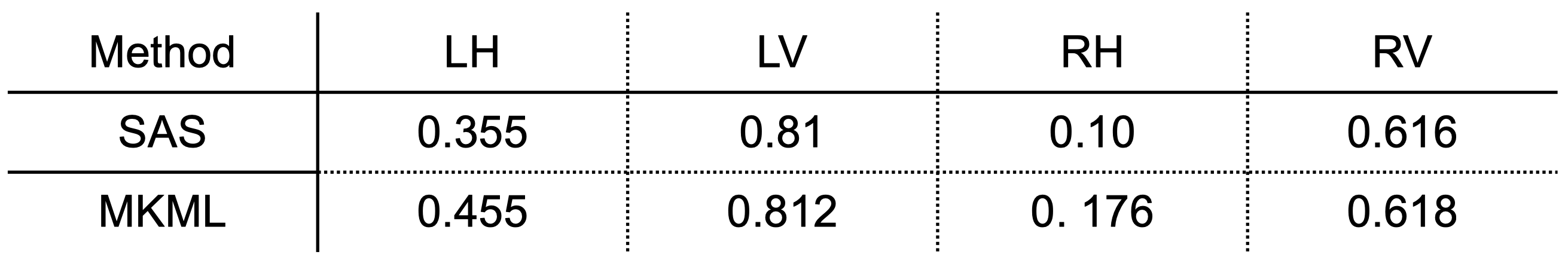}
\vspace{-10pt}
\label{tab:suppTab}
\end{table}

\begin{figure*}[ht]
\centering
\includegraphics[width=13cm]{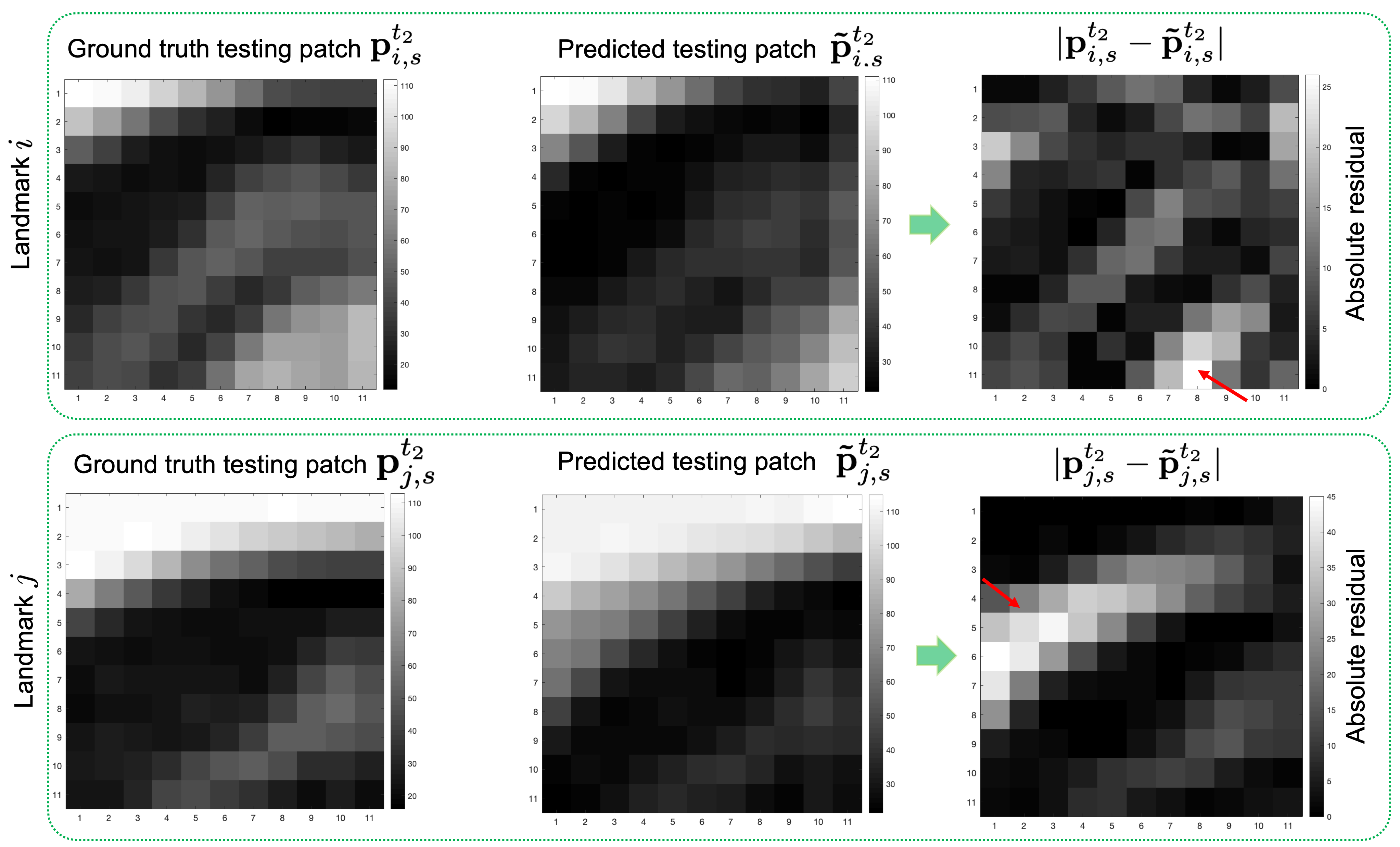}
\caption{\emph{Patch prediction results at two representative landmarks in the left ventricle using supervised atlas selection (SAS) method.} The red arrow points to bright voxels where the residual between the ground truth patch and the predicted patch was large.}
\label{fig:supp3}
\end{figure*}

\textbf{Results and performance.}\footnote{Upon acceptance, the Matlab code of both supervised and unsupervised techniques will be released on GitHub \url{https://github.com/}}. \emph{Prediction}. The mean absolute error in predicted intensity is displayed in \textbf{Fig.}~\ref{fig:4} for each prediction method. The best prediction performance was consistently achieved by SAS in all four ROIs, with a higher prediction error in left and right hippocampi. However, noting that MAE is not a normalized metric, using an alternative normalized evaluation measure can help better discern prediction performance across regions in relation to the classification performance (\textbf{Table}~\ref{tab:1}). Hence, we compute the average Pearson correlation between the ground truth patches and their corresponding predicted patches and report the results in \textbf{Table}~\ref{tab:suppTab}. Clearly, both SAS and MKML produced very promising prediction results in the left and right ventricles, respectively. However, we note a low prediction performance particularly in the right hippocampus. \textbf{Fig.}~\ref{fig:supp3} displays the patch prediction results at two representative landmarks in the left ventricle using (SAS) method. The absolute residual patches show an overall very promising high similarity between the ground truth and predicted patches except a few bright local voxels.

\emph{Classification}. The classification performances of the proposed methods are reported in \textbf{Table}~\ref{tab:1} for each region of interest.  In \textbf{Fig.}~\ref{fig:4}, we compare the performance of the prediction step of our framework side-by-side with the overall classification performance. The best performing method varies by region, but it can be seen that MKML consistently provided improvements over baseline-only classification in three of the four regions. It led to improvements of 5, 10 and 3.33 percentage points in the left lateral ventricle, the left hippocampus and the right hippocampus, respectively. SAS performed less consistently, leading to an improvement over the baseline-only method only in the lateral ventricles. It did however achieve the highest classification accuracy out of all the methods tested on the left lateral ventricle, providing a boost of 8.34 percentage points over the baseline-only method, and gave the highest specificity for the ventricles. 

Ultimately, our results show the most discerning regions for prediction to be in the left hemisphere, and point to the possibility of achieving consistent accuracy improvements upon baseline-only classification using the MKML prediction method proposed in our framework. We would like to emphasise that the main contribution of this work is to highlight the great potential of predicting longitudinal data from baseline data in increasing classification accuracy and its potential use in the diagnosis of neurological disorders. The performance of the strategies proposed can be further improved by leveraging enhancing methods such as learning features using deep learning instead of simply utilizing patch intensities as well as tapping into the active field of feature selection methods.

\section{Discussion}

This paper proposed a joint prediction and classification framework which learns how to predict a subject's MR intensity image at a future timepoint and leverages this prediction to improve diagnosis accuracy \emph{whilst diagnosing based on a single timepoint} using two novel supervised and unsupervised atlas patch selection strategies.

\textbf{Evaluation of results.} The proposed MKML prediction method provides a consistent improvement of up to 10 percentage points in overall classification accuracy in both of the ROIs evaluated in the left hemisphere. The improvement seen in the right hemisphere was limited, with a marginal improvement of 3.33 percentage points in the right lateral ventricle and no improvement in the right hippocampus. It could be the case that the atrophy patterns in the right hemisphere are not discerning enough at the early MCI stage for our framework to detect similarities between eMCI patients and predict the second timepoint as effectively as in the left hemisphere. This could be attributed to the difference in the rates of atrophy that occurs under AD on each side of the brain, as AD-related atrophy is typically more visible in the left hemisphere \cite{Whitwell:2007, Thompson:2003, Karas:2003}.

Supervised atlas selection offered a marginal improvement in overall classification accuracy of 1.67 percentage points in the right lateral ventricle, and offered the best overall classification accuracy of 81.67\% for the left lateral ventricle, though it offered no improvement over the baseline method in the hippocampi in either hemisphere. It also had the highest specificity in both of the lateral ventricles. Such performance can be nicely explained in the light of the results revealed in \textbf{Table}~\ref{tab:suppTab}, where the prediction of the right hippocampus evolution trajectory using SAS was very low compared to other ROIs, thus it did not outperform the baseline method (\textbf{Fig.}~\ref{fig:4}--D). A reason for this could be in fixing the patch size across ROIs, which we opted for in this paper for comparative training at a fixed spatial scale. Also, we note that for SAS the \emph{direction} of the error influences the classification accuracy, as well as the magnitude of the error. For instance, if the true patch has dementia-related atrophy at the voxel being predicted, it could be acceptable to predict an intensity value which is lower than the true voxel intensity, as the predicted voxel intensity should still fall on the same side of the hyperplane of an SVM as the true intensity value and lead to a correct classification, whereas predicting intensities higher than the true voxel value by the same error margin could point to the lack of atrophy and cause a misclassification, despite both predictions having the same mean absolute error.

Remarkably, leveraging the predicted patches by SAS and MKML boosted the classification results --except for the right hippocampus where the prediction error was very large. However, we would like to bring to the attention of the reader that the performance of these methods ultimately depends on the presence of a training atlas patch which is sufficiently similar to the testing patch, since an underlying assumption in both methods is that the best predictors of the closest neighbour of the testing subject will also be good predictors of the testing subject. It is possible, therefore, that these methods could perform more consistently across landmarks with an expanded set of training samples.

\textbf{Limitations and recommendations for future work.} We note that the proposed supervised atlas patch selection strategy assumes that the multiplication of an intensity patch of subject $s$ at baseline by a weighting vector $\mathbf{\alpha}$ can produce the intensity patch of a different subject $s^{\prime}$. By applying the weighting vector $\mathbf{\alpha}$ forward to  selected follow-up atlas patches, we predict the full evolution trajectory of a given testing patch. However, there is no theoretical proof that this assumption holds aside from the fact that the follow-up data predicted in this way generally boosted the subject classification performance as healthy or demented from a single MRI acquired at baseline. This is a proof-of-concept that needs to be further investigated.

Additionally, our current landmark-detection scheme uses threshold values which are pre-defined using the mean minus the standard deviation of intensity distribution in the target ROI. We expect that learning the optimal number $L$ of landmarks from the training dataset and picking the $L$ landmarks with the highest edge density would lead to improved results regardless of the dataset of interest. For instance, one can learn how to identify the best landmarks for the target classification task by leveraging advanced landmark detection techniques such as \cite{Zhang:2016}. We also note that for a relatively fair comparison across regions, subjects and methods, we opted for fixing the patch size (i.e., using a fixed spatial scale), however, the optimal patch size for each region can be alternatively learned. This will be investigated in our future work. Besides, instead of solely relying on the patch intensity for best candidate atlas patch selection, we can use more advanced similarity measures that capture local atrophy (e.g., local deformation field with respect to a healthy template) and structural information (e.g., label patches). Since the proposed framework define the state-of-the-art in joint image evolution trajectory prediction and classification from a single timepoint, we opted for using simple intuitive measures. So far, we have only explored Gaussian kernels \cite{Wang:2017} for learning the patch manifold space. Other manifold learning techniques such as local nonlinear embeddings \cite{Cayton:2005} and Riemannian approaches \cite{Lin:2006} for identifying the best atlas patches for the target prediction and classification tasks.

In our future work, we will evaluate our generic supervised and unsupervised patch-based trajectory evolution prediction and labeling frameworks on various neurological datasets with denser follow-up observations to predict and classify. We will also investigate the effectiveness of the framework on other regions which are highly correlated with AD, such as the entorhinal cortex \cite{Du:2001,vanHoesen:2006}. In addition to predicting image intensity evolution trajectory, one can also extend the proposed strategies to handling 3-dimensional shapes \cite{Fishbaugh:2014,Rekik:2015,Rekik:2017} as well as connectomes \cite{Zhu:2018}, thereby designing a holistic brain evolution trajectory prediction framework from baseline.

\section{Conclusion}
In this paper, we proposed both supervised and unsupervised patch-based evolution trajectory prediction and classification frameworks for accurate diagnosis of early MCI patients. Our initial results show that our framework can consistently boost the classification accuracy by up to 10 percentage points using the MKML method, and by 8 percentage points using the SAS method over the baseline-only method in the left hemisphere, with SAS achieving the highest classification accuracy of 81.67\%. Remarkably, predicting longitudinal MRI data from a single acquisition timepoint largely improved the classification of NC and eMCI subjects, without resorting to any additional enhancing methods (e.g., feature selection techniques). One early measurement proved to be sufficient given that the predictive model works well. In our future work, we intend to further replicate our results in other neurological datasets and incorporate other high-dimensional brain representations such as the cortical surface which atrophies during MCI development. Furthermore, investigating similarity maps quantifying patch similarity between different timepoints \cite{Sanroma:2017} can further improve our classification accuracy while leveraging the predicted follow-up images.

\section*{Acknowledgment}

Data collection and sharing for this project was funded by the Alzheimer's Disease Neuroimaging Initiative (ADNI) (National Institutes of Health Grant U01 AG024904) and DOD ADNI (Department of Defense award number W81XWH-12-2-0012). ADNI is funded by the National Institute on Aging, the National Institute of Biomedical Imaging and Bioengineering, and through generous contributions from the following: AbbVie, Alzheimer's Association; Alzheimer's Drug Discovery Foundation; Araclon Biotech; BioClinica, Inc.; Biogen; Bristol-Myers Squibb Company; CereSpir, Inc.; Cogstate; Eisai Inc.; Elan Pharmaceuticals, Inc.; Eli Lilly and Company; EuroImmun; F. Hoffmann-La Roche Ltd and its affiliated company Genentech, Inc.; Fujirebio; GE Healthcare; IXICO Ltd.; Janssen Alzheimer Immunotherapy Research \& Development, LLC.; Johnson \& Johnson Pharmaceutical Research \& Development LLC.; Lumosity; Lundbeck; Merck \& Co., Inc.; Meso Scale Diagnostics, LLC.; NeuroRx Research; Neurotrack Technologies; Novartis Pharmaceuticals Corporation; Pfizer Inc.; Piramal Imaging; Servier; Takeda Pharmaceutical Company; and Transition Therapeutics. The Canadian Institutes of Health Research is providing funds to support ADNI clinical sites in Canada. Private sector contributions are facilitated by the Foundation for the National Institutes of Health (\url{www.fnih.org}). The grantee organization is the Northern California Institute for Research and Education, and the study is coordinated by the Alzheimer's Therapeutic Research Institute at the University of Southern California. ADNI data are disseminated by the Laboratory for Neuro Imaging at the University of Southern California. 



\ifCLASSOPTIONcaptionsoff
  \newpage
\fi



\bibliography{Biblio}

\begin{thebibliography}{10}
\providecommand{\url}[1]{#1}
\csname url@samestyle\endcsname
\providecommand{\newblock}{\relax}
\providecommand{\bibinfo}[2]{#2}
\providecommand{\BIBentrySTDinterwordspacing}{\spaceskip=0pt\relax}
\providecommand{\BIBentryALTinterwordstretchfactor}{4}
\providecommand{\BIBentryALTinterwordspacing}{\spaceskip=\fontdimen2\font plus
\BIBentryALTinterwordstretchfactor\fontdimen3\font minus
  \fontdimen4\font\relax}
\providecommand{\BIBforeignlanguage}[2]{{%
\expandafter\ifx\csname l@#1\endcsname\relax
\typeout{** WARNING: IEEEtran.bst: No hyphenation pattern has been}%
\typeout{** loaded for the language `#1'. Using the pattern for}%
\typeout{** the default language instead.}%
\else
\language=\csname l@#1\endcsname
\fi
#2}}
\providecommand{\BIBdecl}{\relax}
\BIBdecl

\bibitem{Kalaria:2008}
\BIBentryALTinterwordspacing
R.~N. Kalaria, G.~E. Maestre, R.~Arizaga, R.~P. Friedland, D.~Galasko, K.~Hall,
  J.~A. Luchsinger, A.~Ogunniyi, E.~K. Perry, F.~Potocnik, M.~Prince,
  R.~Stewart, A.~Wimo, Z.-X. Zhang, and P.~Antuono, ``Alzheimer's disease and
  vascular dementia in developing countries: prevalence, management, and risk
  factors,'' \emph{The Lancet Neurology}, vol.~7, no.~9, pp. 812 -- 826, 2008.
  [Online]. Available:
  \url{http://www.sciencedirect.com/science/article/pii/S1474442208701698}
\BIBentrySTDinterwordspacing

\bibitem{Morris:2001}
\BIBentryALTinterwordspacing
M.~JC, S.~M, M.~J, and et~al, ``Mild cognitive impairment represents
  early-stage alzheimer disease,'' \emph{Archives of Neurology}, vol.~58,
  no.~3, pp. 397--405, 2001. [Online]. Available: \url{+
  http://dx.doi.org/10.1001/archneur.58.3.397}
\BIBentrySTDinterwordspacing

\bibitem{Apostolova:2008}
\BIBentryALTinterwordspacing
L.~G. Apostolova and P.~M. Thompson, ``Mapping progressive brain structural
  changes in early alzheimer's disease and mild cognitive impairment,''
  \emph{Neuropsychologia}, vol.~46, no.~6, pp. 1597 -- 1612, 2008, neuroimaging
  of Early Alzheimer's Disease. [Online]. Available:
  \url{http://www.sciencedirect.com/science/article/pii/S0028393207004150}
\BIBentrySTDinterwordspacing

\bibitem{Frisoni:2010}
G.~B. Frisoni, N.~C. Fox, C.~R. Jack, P.~Scheltens, and P.~M. Thompson,
  ``{{T}he clinical use of structural {M}{R}{I} in {A}lzheimer disease},''
  \emph{Nat Rev Neurol}, vol.~6, no.~2, pp. 67--77, Feb 2010.

\bibitem{Coulthard:2017}
\BIBentryALTinterwordspacing
E.~Coulthard and M.~Knight, ``Refining alzheimer's disease diagnosis with
  mri,'' \emph{Brain}, vol. 140, no.~3, pp. 524--526, 2017. [Online].
  Available: \url{http://dx.doi.org/10.1093/brain/aww335}
\BIBentrySTDinterwordspacing

\bibitem{Jessen:2014}
\BIBentryALTinterwordspacing
F.~Jessen, ``Subjective and objective cognitive decline at the pre-dementia
  stage of alzheimer's disease,'' \emph{European Archives of Psychiatry and
  Clinical Neuroscience}, vol. 264, no.~1, pp. 3--7, Nov 2014. [Online].
  Available: \url{https://doi.org/10.1007/s00406-014-0539-z}
\BIBentrySTDinterwordspacing

\bibitem{Magnin:2009}
\BIBentryALTinterwordspacing
B.~Magnin, L.~Mesrob, S.~Kinkingn{\'e}hun, M.~P{\'e}l{\'e}grini-Issac,
  O.~Colliot, M.~Sarazin, B.~Dubois, S.~Leh{\'e}ricy, and H.~Benali, ``Support
  vector machine-based classification of {Alzheimer's disease} from whole-brain
  anatomical mri,'' \emph{Neuroradiology}, vol.~51, no.~2, pp. 73--83, Feb
  2009. [Online]. Available: \url{https://doi.org/10.1007/s00234-008-0463-x}
\BIBentrySTDinterwordspacing

\bibitem{Kloppel:2008}
S.~Kloppel, C.~M. Stonnington, C.~Chu, B.~Draganski, R.~I. Scahill, J.~D.
  Rohrer, N.~C. Fox, C.~R. Jack, J.~Ashburner, and R.~S. Frackowiak,
  ``{{A}utomatic classification of {M}{R} scans in {A}lzheimer's disease},''
  \emph{Brain}, vol. 131, no. Pt 3, pp. 681--689, Mar 2008.

\bibitem{Cuingnet:2011}
\BIBentryALTinterwordspacing
R.~Cuingnet, E.~Gerardin, J.~Tessieras, G.~Auzias, S.~Leh{\'e}ricy, M.-O.
  Habert, M.~Chupin, H.~Benali, and O.~Colliot, ``Automatic classification of
  patients with {Alzheimer's disease from structural MRI}: A comparison of ten
  methods using the {ADNI} database,'' \emph{NeuroImage}, vol.~56, no.~2, pp.
  766 -- 781, 2011, multivariate Decoding and Brain Reading. [Online].
  Available:
  \url{http://www.sciencedirect.com/science/article/pii/S1053811910008578}
\BIBentrySTDinterwordspacing

\bibitem{Sanroma:2017}
G.~Sanroma, V.~Andrea, O.~M. Benkarim, J.~V. Manj{\'o}n, P.~Coup{\'e},
  O.~Camara, G.~Piella, and M.~A. Gonz{\'a}lez~Ballester, ``Early prediction of
  {Alzheimer's Disease} with non-local patch-based longitudinal descriptors,''
  in \emph{Patch-Based Techniques in Medical Imaging}, G.~Wu, B.~C. Munsell,
  Y.~Zhan, W.~Bai, G.~Sanroma, and P.~Coup{\'e}, Eds.\hskip 1em plus 0.5em
  minus 0.4em\relax Cham: Springer International Publishing, 2017, pp. 74--81.

\bibitem{Thung:2016}
\BIBentryALTinterwordspacing
K.-H. Thung, C.-Y. Wee, P.-T. Yap, and D.~Shen, ``Identification of progressive
  mild cognitive impairment patients using incomplete longitudinal mri scans,''
  \emph{Brain Structure and Function}, vol. 221, no.~8, pp. 3979--3995, Nov
  2016. [Online]. Available: \url{https://doi.org/10.1007/s00429-015-1140-6}
\BIBentrySTDinterwordspacing

\bibitem{Zhu:2016}
Y.~Zhu, X.~Zhu, M.~Kim, D.~Shen, and G.~Wu, ``Early diagnosis of {Alzheimer's
  Disease} by joint feature selection and classification on temporally
  structured support vector machine,'' in \emph{Medical Image Computing and
  Computer-Assisted Intervention -- MICCAI 2016}, S.~Ourselin, L.~Joskowicz,
  M.~R. Sabuncu, G.~Unal, and W.~Wells, Eds.\hskip 1em plus 0.5em minus
  0.4em\relax Cham: Springer International Publishing, 2016, pp. 264--272.

\bibitem{Zhang:2012}
\BIBentryALTinterwordspacing
D.~Zhang, D.~Shen, and A.~D.~N. Initiative, ``Predicting future clinical
  changes of mci patients using longitudinal and multimodal biomarkers,''
  \emph{PLOS ONE}, vol.~7, no.~3, pp. 1--15, 03 2012. [Online]. Available:
  \url{https://doi.org/10.1371/journal.pone.0033182}
\BIBentrySTDinterwordspacing

\bibitem{Maliszewska:2017}
E.~Maliszewska-Cyna, M.~Lynch, J.~Jordan~Oore, P.~Michael~Nagy, and I.~Aubert,
  ``The benefits of exercise and metabolic interventions for the prevention and
  early treatment of {Alzheimer's disease},'' \emph{Current Alzheimer
  Research}, vol.~14, no.~1, pp. 47--60, 2017.

\bibitem{Soussia:2018a}
M.~Soussia and I.~Rekik, ``A review on image-and network-based brain data
  analysis techniques for {Alzheimer's Disease} diagnosis reveals a gap in
  developing predictive methods for prognosis,'' \emph{arXiv preprint
  arXiv:1808.01951}, 2018.

\bibitem{Rekik:2015}
I.~Rekik, G.~Li, W.~Lin, and D.~Shen, ``Prediction of longitudinal development
  of infant cortical surface shape using a {4D} current-based learning
  framework,'' \emph{International Conference on Information Processing in
  Medical Imaging}, pp. 576--587, 2015.

\bibitem{Rekik:2016a}
------, ``Predicting infant cortical surface development using a {4D}
  varifold-based learning framework and local topography-based shape
  morphing,'' \emph{Medical image analysis}, vol.~28, pp. 1--12, 2016.

\bibitem{Rekik:2017}
I.~Rekik, G.~Li, P.-T. Yap, G.~Chen, W.~Lin, and D.~Shen, ``Joint prediction of
  longitudinal development of cortical surfaces and white matter fibers from
  {neonatal MRI},'' \emph{NeuroImage}, vol. 152, pp. 411--424, 2017.

\bibitem{Sanroma:2014}
G.~Sanroma, G.~Wu, Y.~Gao, and D.~Shen, ``Learning-based atlas selection for
  multiple-atlas segmentation,'' \emph{Proceedings of the IEEE Conference on
  Computer Vision and Pattern Recognition}, pp. 3111--3117, 2014.

\bibitem{Wang:2017}
\BIBentryALTinterwordspacing
B.~Wang, J.~Zhu, E.~Pierson, D.~Ramazzotti, and S.~Batzoglou, ``Visualization
  and analysis of single-cell rna-seq data by kernel-based similarity
  learning,'' \emph{Nature Methods}, vol.~14, pp. 414 EP --, Mar 2017.
  [Online]. Available: \url{http://dx.doi.org/10.1038/nmeth.4207}
\BIBentrySTDinterwordspacing

\bibitem{Gonen:2011}
M.~G{\"o}nen and E.~Alpayd{\i}n, ``Multiple kernel learning algorithms,''
  \emph{Journal of machine learning research}, vol.~12, no. Jul, pp.
  2211--2268, 2011.

\bibitem{Mueller:2005}
S.~G. Mueller, M.~W. Weiner, L.~J. Thal, R.~C. Petersen, C.~Jack, W.~Jagust,
  J.~Q. Trojanowski, A.~W. Toga, and L.~Beckett, ``The {Alzheimer's} disease
  neuroimaging initiative,'' \emph{Neuroimaging Clinics}, vol.~15, no.~4, pp.
  869--877, 2005.

\bibitem{Jack:2008}
C.~R. Jack~Jr, M.~A. Bernstein, N.~C. Fox, P.~Thompson, G.~Alexander,
  D.~Harvey, B.~Borowski, P.~J. Britson, J.~L.~Whitwell, C.~Ward \emph{et~al.},
  ``The {Alzheimer's disease neuroimaging initiative (ADNI): MRI methods},''
  \emph{Journal of Magnetic Resonance Imaging: An Official Journal of the
  International Society for Magnetic Resonance in Medicine}, vol.~27, no.~4,
  pp. 685--691, 2008.

\bibitem{Aisen:2010}
P.~S. Aisen, R.~C. Petersen, M.~C. Donohue, A.~Gamst, R.~Raman, R.~G. Thomas,
  S.~Walter, J.~Q. Trojanowski, L.~M. Shaw, L.~A. Beckett \emph{et~al.},
  ``{Clinical Core of the Alzheimer's Disease Neuroimaging Initiative: progress
  and plans},'' \emph{Alzheimer's \& Dementia}, vol.~6, no.~3, pp. 239--246,
  2010.

\bibitem{Tan:2013}
E.~T. Tan, L.~Marinelli, Z.~W. Slavens, K.~F. King, and C.~J. Hardy, ``Improved
  correction for gradient nonlinearity effects in diffusion-weighted imaging,''
  \emph{Journal of Magnetic Resonance Imaging}, vol.~38, no.~2, pp. 448--453,
  2013.

\bibitem{Boyes:2008}
R.~G. Boyes, J.~L. Gunter, C.~Frost, A.~L. Janke, T.~Yeatman, D.~L. Hill, M.~A.
  Bernstein, P.~M. Thompson, M.~W. Weiner, N.~Schuff \emph{et~al.}, ``Intensity
  non-uniformity correction using n3 on 3-t scanners with multichannel phased
  array coils,'' \emph{Neuroimage}, vol.~39, no.~4, pp. 1752--1762, 2008.

\bibitem{Fischl:2012}
B.~Fischl, ``Freesurfer,'' \emph{Neuroimage}, vol.~62, no.~2, pp. 774--781,
  2012.

\bibitem{Collins:1995}
D.~L. Collins, C.~J. Holmes, T.~M. Peters, and A.~C. Evans, ``Automatic {3-D}
  model-based neuroanatomical segmentation,'' \emph{Human brain mapping},
  vol.~3, no.~3, pp. 190--208, 1995.

\bibitem{Morra:2009}
\BIBentryALTinterwordspacing
J.~H. Morra, Z.~Tu, L.~G. Apostolova, A.~E. Green, C.~Avedissian, S.~K. Madsen,
  N.~Parikshak, X.~Hua, A.~W. Toga, C.~R. Jack, N.~Schuff, M.~W. Weiner, and
  P.~M. Thompson, ``Automated 3d mapping of hippocampal atrophy and its
  clinical correlates in 400 subjects with alzheimer's disease, mild cognitive
  impairment, and elderly controls,'' \emph{Human Brain Mapping}, vol.~30,
  no.~9, pp. 2766--2788, 2009. [Online]. Available:
  \url{https://onlinelibrary.wiley.com/doi/abs/10.1002/hbm.20708}
\BIBentrySTDinterwordspacing

\bibitem{Apostolova:2006}
\BIBentryALTinterwordspacing
L.~G. Apostolova, I.~D. Dinov, R.~A. Dutton, K.~M. Hayashi, A.~W. Toga, J.~L.
  Cummings, and P.~M. Thompson, ``3d comparison of hippocampal atrophy in
  amnestic mild cognitive impairment and alzheimer's disease,'' \emph{Brain},
  vol. 129, no.~11, pp. 2867--2873, 2006. [Online]. Available:
  \url{http://dx.doi.org/10.1093/brain/awl274}
\BIBentrySTDinterwordspacing

\bibitem{Deweer:1995}
\BIBentryALTinterwordspacing
B.~Deweer, S.~Leh{\'e}ricy, B.~Pillon, M.~Baulac, J.~Chiras, C.~Marsault,
  Y.~Agid, and B.~Dubois, ``Memory disorders in probable
  alzheimer{\textquoteright}s disease: the role of hippocampal atrophy as shown
  with mri.'' \emph{Journal of Neurology, Neurosurgery \& Psychiatry}, vol.~58,
  no.~5, pp. 590--597, 1995. [Online]. Available:
  \url{http://jnnp.bmj.com/content/58/5/590}
\BIBentrySTDinterwordspacing

\bibitem{deLeon:1993}
\BIBentryALTinterwordspacing
M.~J. de~Leon, J.~Golomb, A.~E. George, A.~Convit, C.~Y. Tarshish, T.~McRae,
  S.~De~Santi, G.~Smith, S.~H. Ferris, and M.~Noz, ``The radiologic prediction
  of alzheimer disease: the atrophic hippocampal formation.'' \emph{American
  Journal of Neuroradiology}, vol.~14, no.~4, pp. 897--906, 1993. [Online].
  Available: \url{http://www.ajnr.org/content/14/4/897}
\BIBentrySTDinterwordspacing

\bibitem{Thompson:2004}
\BIBentryALTinterwordspacing
P.~M. Thompson, K.~M. Hayashi, G.~I. de~Zubicaray, A.~L. Janke, S.~E. Rose,
  J.~Semple, M.~S. Hong, D.~H. Herman, D.~Gravano, D.~M. Doddrell, and A.~W.
  Toga, ``Mapping hippocampal and ventricular change in alzheimer disease,''
  \emph{NeuroImage}, vol.~22, no.~4, pp. 1754 -- 1766, 2004. [Online].
  Available:
  \url{http://www.sciencedirect.com/science/article/pii/S105381190400196X}
\BIBentrySTDinterwordspacing

\bibitem{Nestor:2008}
\BIBentryALTinterwordspacing
S.~M. Nestor, R.~Rupsingh, M.~Borrie, M.~Smith, V.~Accomazzi, J.~L. Wells,
  J.~Fogarty, R.~Bartha, and the Alzheimer's Disease Neuroimaging~Initiative,
  ``Ventricular enlargement as a possible measure of alzheimer's disease
  progression validated using the alzheimer's disease neuroimaging initiative
  database,'' \emph{Brain}, vol. 131, no.~9, pp. 2443--2454, 2008. [Online].
  Available: \url{http://dx.doi.org/10.1093/brain/awn146}
\BIBentrySTDinterwordspacing

\bibitem{Whitwell:2007}
\BIBentryALTinterwordspacing
J.~L. Whitwell, S.~A. Przybelski, S.~D. Weigand, D.~S. Knopman, B.~F. Boeve,
  R.~C. Petersen, and C.~R. Jack, Jr, ``3d maps from multiple mri illustrate
  changing atrophy patterns as subjects progress from mild cognitive impairment
  to alzheimer's disease,'' \emph{Brain}, vol. 130, no.~7, pp. 1777--1786,
  2007. [Online]. Available: \url{http://dx.doi.org/10.1093/brain/awm112}
\BIBentrySTDinterwordspacing

\bibitem{Thompson:2003}
P.~M. Thompson, K.~M. Hayashi, G.~de~Zubicaray, A.~L. Janke, S.~E. Rose,
  J.~Semple, D.~Herman, M.~S. Hong, S.~S. Dittmer, D.~M. Doddrell, and A.~W.
  Toga, ``{Dynamics of gray matter loss in {A}lzheimer's disease},'' \emph{J.
  Neurosci.}, vol.~23, no.~3, pp. 994--1005, Feb 2003.

\bibitem{Karas:2003}
\BIBentryALTinterwordspacing
G.~Karas, E.~Burton, S.~Rombouts, R.~van Schijndel, J.~O'Brien, P.~Scheltens,
  I.~McKeith, D.~Williams, C.~Ballard, and F.~Barkhof, ``A comprehensive study
  of gray matter loss in patients with alzheimer's disease using optimized
  voxel-based morphometry,'' \emph{NeuroImage}, vol.~18, no.~4, pp. 895 -- 907,
  2003. [Online]. Available:
  \url{http://www.sciencedirect.com/science/article/pii/S1053811903000417}
\BIBentrySTDinterwordspacing

\bibitem{Zhang:2016}
J.~Zhang, Y.~Gao, Y.~Gao, B.~C. Munsell, and D.~Shen, ``Detecting anatomical
  landmarks for fast {Alzheimer's} disease diagnosis,'' \emph{IEEE transactions
  on medical imaging}, vol.~35, no.~12, pp. 2524--2533, 2016.

\bibitem{Cayton:2005}
L.~Cayton, ``Algorithms for manifold learning,'' \emph{Univ. of California at
  San Diego Tech. Rep}, vol.~12, no. 1-17, p.~1, 2005.

\bibitem{Lin:2006}
T.~Lin, H.~Zha, and S.~U. Lee, ``Riemannian manifold learning for nonlinear
  dimensionality reduction,'' pp. 44--55, 2006.

\bibitem{Du:2001}
\BIBentryALTinterwordspacing
A.~T. Du, N.~Schuff, D.~Amend, M.~P. Laakso, Y.~Y. Hsu, W.~J. Jagust, K.~Yaffe,
  J.~H. Kramer, B.~Reed, D.~Norman, H.~C. Chui, and M.~W. Weiner, ``Magnetic
  resonance imaging of the entorhinal cortex and hippocampus in mild cognitive
  impairment and alzheimer{\textquoteright}s disease,'' \emph{Journal of
  Neurology, Neurosurgery \& Psychiatry}, vol.~71, no.~4, pp. 441--447, 2001.
  [Online]. Available: \url{http://jnnp.bmj.com/content/71/4/441}
\BIBentrySTDinterwordspacing

\bibitem{vanHoesen:2006}
G.~W.~V. Hoesen, J.~C. Augustinack, J.~Dierking, S.~J. Redman, and
  R.~Thangavel, ``The parahippocampal gyrus in alzheimer's disease: Clinical
  and preclinical neuroanatomical correlates,'' \emph{Annals of the New York
  Academy of Sciences}, vol. 911, no.~1, pp. 254--274, 2006.

\bibitem{Fishbaugh:2014}
J.~Fishbaugh, M.~Prastawa, G.~Gerig, and S.~Durrleman, ``Geodesic regression of
  image and shape data for improved modeling of 4d trajectories,''
  \emph{Biomedical Imaging (ISBI), 2014 IEEE 11th International Symposium on},
  pp. 385--388, 2014.

\bibitem{Zhu:2018}
M.~Zhu and I.~Rekik, ``Multi-view brain network prediction from a source view
  using sample selection via cca-based multi-kernel connectomic manifold
  learning,'' \emph{International Workshop on PRedictive Intelligence In
  MEdicine}, pp. 94--102, 2018.

\end{thebibliography}
\bibliographystyle{IEEEtran}

%







\end{document}